\newif\ifisdraft\isdrafttrue
\isdraftfalse

\newif\ifusechanges\usechangesfalse

\ifusechanges\isdrafttrue\fi

\ifisdraft
    \documentclass[preprint,prl,amsmath,amssymb,nofootinbib,superscriptaddress]{revtex4}
\else
    \documentclass[twocolumn,prl,amsmath,amssymb,nofootinbib,superscriptaddress]{revtex4}
\fi

\usepackage{graphicx}                  
\usepackage{mathrsfs}                  

\ifusechanges
   \usepackage[ulem = normalem]{changes}
\else
   \usepackage[normalem]{ulem}
   \usepackage{color}
\fi

\begin{document}

\newcommand{\bvec}[1]{\ensuremath{\mathbf{#1}}}

\newcommand{\fig}[1]{Fig.~\ref{#1}}
\newcommand{\eqn}[1]{Eqn.~(\ref{#1})}
\newcommand{\tableref}[1]{Table~\ref{#1}}

\newcommand{\tab}{\hspace{3ex}}
\newcommand{\noi}{\noindent}

\setlength{\tabcolsep}{2 ex} 
\newcommand{\mctwo}[1]{\multicolumn{2}{c}{#1}}

\newenvironment
	{revtabular}[1]
	{\begin{tabular}{#1} \toprule }
	{\\ \botrule \end{tabular}}

\graphicspath{.\figures}
\newdimen\figwidth \figwidth=3.3in

\ifusechanges
   \definechangesauthor[Brahms]{NB}{red}
   \definechangesauthor[Walker]{TW}{cyan}
   \definechangesauthor[Tscherbul]{TT}{purple}
   \definechangesauthor[Doyle]{JD}{green}
   \definechangesauthor[Klos]{JK}{orange}
   \definechangesauthor[Dalgarno]{AD}{brown}
   \definechangesauthor[Sadeghpour]{HS}{magenta}
   \newcommand{\remark}[2][]{\added[#1][#2]{}}
   \newcommand{\inlineremark}[2][]{\ifisdraft \added[#1]{#2} \fi}
\else
   \newcommand{\added}[2][]
       {\ifisdraft\textbf{#2}\textsuperscript{#1}\else{#2}\fi}
   \newcommand{\deleted}[2][]
       {\ifisdraft\sout{#2}\textsuperscript{#1}\fi}
   \newcommand{\replaced}[3][]
       {\ifisdraft\textbf{#2}\sout{#3}\textsuperscript{#1}\else{#2}\fi}
   \newcommand{\remark}[2][]{\ifisdraft\textsuperscript{#1}\footnote{#2}\fi}
   \newcommand{\inlineremark}[2][]{\ifisdraft\added[#1]{#2}\fi}
\fi


\renewcommand{\k}{\ensuremath{\kappa}}		
\newcommand{\K}{\ensuremath{K}}				
\newcommand{\D}{\ensuremath{D}}				
\newcommand{\z}{\ensuremath{\zeta}}			

\newcommand{\Nnorm}{\ensuremath{N}}
\newcommand{\EN}{\ensuremath{E_{N}}}

\newcommand{\T}{\ensuremath{T}}

\newcommand{\mathhe}{\mathrm{He}}
\newcommand{\xhe}{\ensuremath{X\mathrm{He}}}
\newcommand{\nhe}{\ensuremath{n_\mathrm{He}}}
\newcommand{\nxhe}{\ensuremath{n_{\xhe}}}
\newcommand{\nx}{\ensuremath{n_X}}
\newcommand{\tf}{\ensuremath{\tau_\mathrm{f}}}
\newcommand{\td}{\ensuremath{\tau_\mathrm{d}}}
\newcommand{\tc}{\ensuremath{\tau_\mathrm{c}}}
\newcommand{\he}{\textrm{He}}
\newcommand{\hethree}{\ensuremath{^3\mathrm{He}}}
\newcommand{\aghe}{\textrm{Ag}\he}
\newcommand{\aghethree}{\textrm{Ag}\hethree}
\newcommand{\aghecol}{Ag--\hethree}
\newcommand{\aghehecol}{\aghethree--\hethree}
\newcommand{\vdw}{van der Waals molecule}
\newcommand{\vbar}{\ensuremath{\bar v}}

\def\longrightharpoonup{\relbar\joinrel\rightharpoonup}
\def\longleftharpoondown{\leftharpoondown\joinrel\relbar}

\def\longrightleftharpoons{
  \mathop{
    \vcenter{
      \hbox{
      \ooalign{
        \raise1pt\hbox{$\longrightharpoonup\joinrel$}\crcr
	  \lower1pt\hbox{$\longleftharpoondown\joinrel$}
	  }
      }
    }
  }
}

\newcommand{\rates}[2]{\displaystyle
  \mathrel{\longrightleftharpoons^{#1\mathstrut}_{#2}}}

\newcommand{\Thad}[1]{\added[TW]{#1}}


\newcommand{\enAgHeGround}{\ensuremath{-1.40}}			
\newcommand{\BAgHeGround}{\ensuremath{0.176}}			
\newcommand{\cmone}{\ensuremath{\:\mathrm{cm}^{-1}}}	
\newcommand{\cmthree}{\ensuremath{\mathrm{cm}^{-3}}}
\newcommand{\cmthrees}{\ensuremath{\mathrm{cm}^3/\mathrm{s}}}

\newcommand{\massaghe}{\ensuremath{2.92\:\mathrm{amu}}}	

\newcommand{\maxmolfrac}{$14\%$}			

\newcommand{\molconfig}{$X{^2\Sigma}^{+}$}	
\newcommand{\excitedconfig}{$A'{^2\Pi}_{3/2}$}
\newcommand{\moldipole}{0.04 debye}		
\newcommand{\maxmolnumber}{$2\times 10^{12}$}

\newcommand{\savalue}{\ensuremath{3.4\times 10^{-14}\:\mathrm{cm}^2}}
\newcommand{\gmvalue}{\ensuremath{1.85\times 10^{-5}}}

\title{Formation  of van der Waals molecules in buffer gas cooled magnetic traps}

\author{N.\ Brahms}
\affiliation{Department of Physics, University of California, Berkeley, California  97420}
\affiliation{Harvard-MIT Center for Ultracold Atoms, Cambridge, Massachusetts 02138}
\author{T.\ V.\ Tscherbul}
\affiliation{Harvard-MIT Center for Ultracold Atoms, Cambridge, Massachusetts 02138}
\affiliation{ITAMP, Harvard-Smithsonian Center for Astrophysics, Cambridge, Massachusetts 02138}
\author{P. Zhang}
\affiliation{ITAMP, Harvard-Smithsonian Center for Astrophysics, Cambridge, Massachusetts 02138}
\author{J.\ K\l os}
\affiliation{Department of Chemistry and Biochemistry, University of Maryland, College Park, Maryland 20742} 
\author{H.\ R.\ Sadeghpour}
\affiliation{ITAMP, Harvard-Smithsonian Center for Astrophysics, Cambridge, Massachusetts 02138}
\author{A.\ Dalgarno}
\affiliation{Harvard-MIT Center for Ultracold Atoms, Cambridge, Massachusetts 02138}
\affiliation{ITAMP, Harvard-Smithsonian Center for Astrophysics, Cambridge, Massachusetts 02138}
\author{J.\ M.\ Doyle}
\affiliation{Harvard-MIT Center for Ultracold Atoms, Cambridge, Massachusetts 02138}
\affiliation{Department of Physics, Harvard University, Cambridge, Massachusetts 02138}
\author{T.\ G.\ Walker}
\affiliation{Department of Physics, University of Wisconsin-Madison, Madison, Wisconsin  53715}
\date{\today}

\begin{abstract}
{We show that a large class of helium-containing cold polar molecules form readily in a cryogenic buffer gas, achieving densities as high as $10^{12}~\mathrm{cm}^{-3}$.  We explore the spin relaxation of these molecules in buffer gas loaded magnetic traps, and identify a loss mechanism based on Landau-Zener transitions arising from the anisotropic hyperfine interaction.  
Our results show that the recently observed strong $T^6$ thermal dependence of spin change in buffer gas trapped silver (Ag) is accounted for by the formation and spin change of \aghethree, thus providing evidence for molecular formation in a buffer gas trap.}
\end{abstract}

\maketitle




Techniques to produce, trap, and manipulate cold neutral and ionic molecules hold great promise for new discoveries within chemistry and physics.  Cold molecules offer new ways to improve precision spectroscopy and measurement \cite{ColdMoleculesBook}, and \added{explore} quantum collective phenomena such as quantum magnetism \cite{QuantumMagnetism} and  high-$T_c$ superconductivity \cite{SpinExchange}. The long-range anisotropic interactions of polar molecules in optical lattices might be tuned with electric fields, allowing for the design of robust quantum information processing \cite{demille:prl:polarqi} and quantum simulations of condensed-matter systems \cite{Bloch}.

To date, two types of cold trapped molecules have been produced in the laboratory.  The first are molecules formed via photo- and magneto-association from ultracold atomic gases, creating ultracold alkali dimers \cite{demille:prl:rbcs,jin:natphys:krb} and trimers \cite{grimm:nature:efimovtrimer}.  The second type are stable dipolar molecules slowed or cooled from high temperatures using Stark deceleration \cite{meijer:prl:decel} or buffer gas cooling \cite{doyle:prl:nhtrap}.  Here we introduce a third class of trappable molecules---van der Waals (vdW) complexes---and provide compelling theoretical support for the formation of \aghethree\ heteronuclear molecules in a recent experiment, in which Ag atoms were magnetically trapped in the presence of a dense \hethree\ gas \cite{brahms:agTrapping}.

Weakly bound vdW complexes play a key role in chemical reaction dynamics, surface interactions, and non-linear optical phenomena in dense atomic and molecular vapors \cite{levy:vdw:review,hutson:vdw:intermolecularForces:review,stoll:he2,stoll:he3,vilesov:clusters:superfluidity}. In particular, vdW complexes containing He atoms have attracted special attention due to the existence of quantum halos in $^4$He$_2$ \cite{stoll:he2} and similar exotic many-body states in $^4$He$_3$  \cite{stoll:he3}. 
Due to their small binding energies and vulnerability to collisions, He-containing complexes have previously only been observed in the collision-free, transient conditions of supersonic jets \cite{levy:vdw:review}.

Here we show how a wide variety of cold, trappable vdW molecules can be produced in buffer gas loaded magnetic traps.  We provide a general model for the formation of vdW pairs and calculate expected molecular densities for a sampling of magnetically trappable species.  Using \textit{ab initio} calculations of interaction potentials and hyperfine interactions, quantum calculations of collision dynamics, and detailed Monte Carlo simulations, we show that recent spin relaxation measurements \cite{brahms:agTrapping} provide convincing evidence for the formation of \aghe\ molecules.  
\\

\noindent\textit{Molecular formation kinetics.}
We begin by outlining the general model for the formation of $X$He molecules via three-body collisions of a species $X$ (atoms, molecules, or ions) with He atoms:
\begin{equation}\label{3BR}
X + \mathhe + \mathhe \rates{K}{D}  X\mathhe + \mathhe,
\end{equation}
where $K$ and $D$ are the rate constants for three-body recombination and collision-induced dissociation.
From this equation, we obtain the formation and dissociation kinetics:
\begin{equation}\label{molkin}
\dot{n}_{\xhe} = -\dot{n}_X = n_X/\tf -n_{\xhe}/\td,
\end{equation}
where $n_i$ denotes the density of species $i$, and the formation time $\tf$ and dissociation time $\td$ obey $1/\tf=K\nhe^2$ and $1/\td=D\,\nhe$.

In thermal equilibrium, $\dot{n}_{\xhe} = 0$ implies $n_{\xhe} = \k(T) \nx \nhe$,  where the chemical equilibrium coefficient $\k(T) = K/D$ can be evaluated from statistical thermodynamics \cite{reif}:
\begin{equation}
\k = \frac{\nxhe}{\nx \nhe} = \left ( \frac{h^2}{2\pi \, \mu \, k_B \, T} \right )^{3/2}\, \sum_i g_i e^{-\epsilon_i/ k_B\, T},
\label{eqn:kappa:def}
\end{equation}
where $k_B$ is Boltzmann's constant, $T$ is \added[AD]{the} temperature, $\mu$ is the reduced mass of the $\xhe$ molecule, and $-\epsilon_i$ is the dissociation energy of molecular state $i$, with degeneracy $g_i$.  \eqn{eqn:kappa:def} \added[AD]{shows} that $\nxhe$ increases exponentially as the temperature is lowered, or as the binding energy of the molecule is increased.  A large number of vdW complexes have binding energies that are comparable to or greater than 1~K, and thus formation is thermodynamically favored in buffer gas cooling experiments.  Table~\ref{table:dimerEnergies} gives a sampling of candidate species for molecular production and predicted population ratios $\frac{\nxhe}{\nx} = \k\nhe$ at standard temperatures and buffer gas densities.
\begin{table}
\setcounter{footnote}{1}
\caption{Predicted ground-state energies $\epsilon_0$ and molecular population ratios $\frac{\nxhe}{\nx} = \k\nhe$ of a few species compatible with buffer gas loaded magnetic traps \cite{dimerTable1,dimerTable2,dimerTable3}.} 
\label{table:dimerEnergies}
\footnotetext[1]{
Energies in cm$^{-1}$, $1~\mathrm{cm}^{-1}\approx 1.4~\mathrm{K}$.  0 is the dissociation energy.}
\footnotetext[2]{
Population ratios are evaluated for $n_\text{He} = 3\times 10^{16}~\mathrm{cm}^{-3}$, at 300~mK for $X^3$He and at 700~mK for $X^4$He molecules, using only the level with energy $\epsilon_0$.}
\begin{revtabular}{ccr@{}lr@{}lr@{}lr@{}l}
     &       & \multicolumn{4}{c}{$X$$^3$He} & \multicolumn{4}{c}{$X$$^4$He} \\
Atom & State & \mctwo{$-\epsilon_0$\footnotemark[1]} & \mctwo{$\frac{\nxhe}{\nx}$\footnotemark[2]} & \mctwo{$-\epsilon_0$\footnotemark[1]} & \mctwo{$\frac{\nxhe}{\nx}$\footnotemark[2]} \\
\colrule
N	&	$^4S_{3/2}$	&	2&.13	&	5&.4		&	2&.85	&	2&.3	\\
P	&	$^4S_{3/2}$	&	2&.70	& 84&			&	3&.42	&	18&	\\
Cu	&	$^2S_{1/2}$	&	0&.90	&	0&.015	&	1&.26	&	0&.008	\\
Ag	&  $^2S_{1/2}$	&	1&.40	&	0&.16		&	1&.85	&	0&.06	\\
Au	&	$^2S_{1/2}$	&	4&.91	&	\mctwo{$3\times 10^6$}	&	5&.87	&	\mctwo{$1\times 10^5$}
\end{revtabular}
\end{table}

The molecular density will come into equilibrium on a timescale \tc, determined by \tf\ and \td.  From \eqn{molkin}:
\begin{equation}
\tau_c^{-1}=\td^{-1}+\tf^{-1}=D\,\nhe\,(1+\k\nhe).
\end{equation}
An estimate for the rate of approach to chemical equilibrium is obtained by assuming that $D\sim\sigma v e^{\epsilon/k_BT}$, with $\sigma$ a typical low-temperature gas kinetic cross section and the exponential factor giving the relative probability that a colliding He atom has enough energy to dissociate the molecule.
Choosing worst-case values $\sigma=10^{-15}$~cm$^2$, $v=50$~m/s, $T=0.3$~K, and $n_\text{He}=10^{16}$~cm$^{-3}$ gives $\tc < 20$~ms for species with $|\epsilon_0 |< 1.4~\mathrm{cm}^{-1}$, and less than 400~ms for all $\epsilon_0$.  Typical buffer gas trap lifetimes of seconds, therefore, indicate that the molecule population reaches equilibrium even for those species having $|\epsilon |/k_B$ larger than 1~K.
\\
\\
\noindent\textit{Spin stability.}
We investigate the spin stability of magnetically trapped molecules, considering those formed from $S$--state atoms for simplicity. The Hamiltonian for the ground electronic and vibrational state of an $X$He($\Sigma$) molecule in a magnetic field of strength $B$ is
\added[NB]{\begin{multline}\label{eqn:HAtomic}
\hat{H}_\text{mol} = \epsilon_{N}
  + A_X\bvec{I}_X\cdot \bvec{S} + \bvec{B}\cdot\left (2\mu_B\bvec{S} + \mu_X\bvec{I}_X + \mu_\mathrm{He}\bvec{I}_{\mathrm{He}} \right ) + \\ \gamma\bvec{N}\cdot \bvec{S} + A_\mathrm{He}\bvec{I}_\mathrm{He} \cdot \bvec{S}  + c \sqrt{\frac{8\pi}{15}} \sum_{q=-2} ^2 Y^*_{2,q}(\hat{r})[\bvec{I}_\mathrm{He}\otimes\bvec{S}]^{(2)}_q,
\end{multline}
where $\epsilon_{N}$ are the rotational energy levels, $\bvec{N}$ is the rotational angular momentum, $\bvec{S}$ is the electron spin, $\gamma$ is the spin-rotation constant, $\bvec{I}_X$ is the nuclear spin of $X$ with moment $\mu_X$, $A_X$ is the atomic hyperfine constant, and $\mu_B$ is the Bohr magneton.  The last two terms in \eqn{eqn:HAtomic} describe the isotropic and anisotropic hyperfine interaction of $\bvec{S}$ with a $^3$He nuclear spin $\bvec{I}_\mathrm{He}$.  We neglect the weak anisotropic component of the $\bvec{I}_X$--$\bvec{S}$ interaction.}

The typical trap used for buffer gas cooling is a magnetic quadrupole, with $|B|=0$ at the center, linearly increasing to a few Teslas at the edge.  
The weak-field-seeking states of both the atoms and the molecules are stably confined in such a trap.

\added[NB]{Collisions with free He atoms can cause loss of the trapped \xhe\ molecules \cite{bouchiat:alkalimolecules}.  The coherence of the precessing electronic spin is lost in the collision, and the internal $\bvec{N}$--$\bvec{S}$ and $\bvec{I}_\mathrm{He}$--$\bvec{S}$ interactions can cause spin relaxation, with the Zeeman energy released into the kinetic energy of the collision complex.  However, the $B$ field strongly decouples the electron spin from other internal angular momenta.  The probability to suffer spin relaxation in a collision due to an interaction of the form $a \bvec{J}\cdot\bvec{S}$, after averaging over the trap distribution of magnetic fields, is
$p = \frac{a^2\langle J (J+1)\rangle}{24\,(k_BT)^2}$.
Estimates of $\gamma$, $A_\mathrm{He}$, and $c$, all on the order of a few MHz \cite{walker:pra1997:spinRotation}, give $p\sim 10^{-9}$, far too small to be a significant factor for trap lifetimes.}

\begin{figure}
\includegraphics[width=\figwidth]{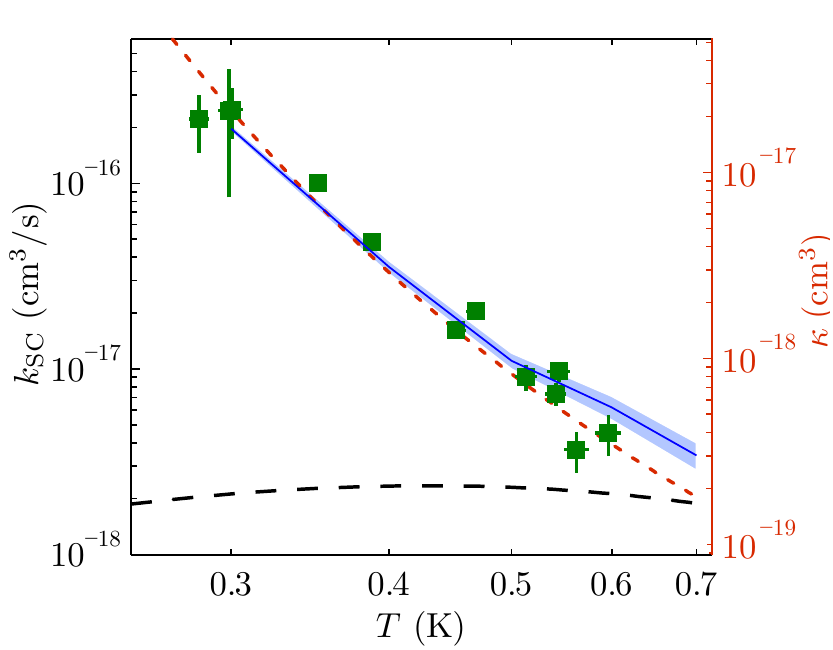}
\caption{Experimentally observed \aghecol\ spin change rate coefficients (squares), compared with the atomic spin-relaxation model (dashed line), and molecular spin change rate coefficients, obtained from Monte Carlo simulations (solid line, shaded area indicates 68\% C.I.).  The simulations were used to fit for the \aghethree\ binding energy at the 380~mK datum, yielding $\epsilon_0 = -1.53\pm 0.08~\mathrm{cm}^{-1}$.  Also shown is a comparison to the chemical equilibrium coefficient $\kappa(T)$ (dotted line, right axis).
}
\label{fig:agHeRate}
\end{figure}

We have identified a spin-relaxation mechanism, due to adiabatic transitions at avoided level crossings induced by the anisotropic hyperfine interaction, explained in depth in our discussion of \aghe\ below.  While this is the dominant loss process for certain $X\hethree$ pairs, we find that trap lifetimes $\gtrsim 1$~s are still achievable for $\Sigma$--state vdW heteronuclear molecules formed in buffer gas traps.
\\
\\
\noindent\textit{Ag$^3\!$He molecules}.  We will now sketch our analysis for the case of \aghethree.  Recently, we reported on the trapping of Ag atoms in a \hethree\ buffer gas \cite{brahms:agTrapping}.  The observed Ag spin-change rate decreased by almost two orders of magnitude as temperature was increased from 300 to 700~mK; the data are reproduced in \fig{fig:agHeRate}.

\begin{figure*}
  \begin{minipage}[t]{2.30in}
       {\includegraphics[width=2.3in]{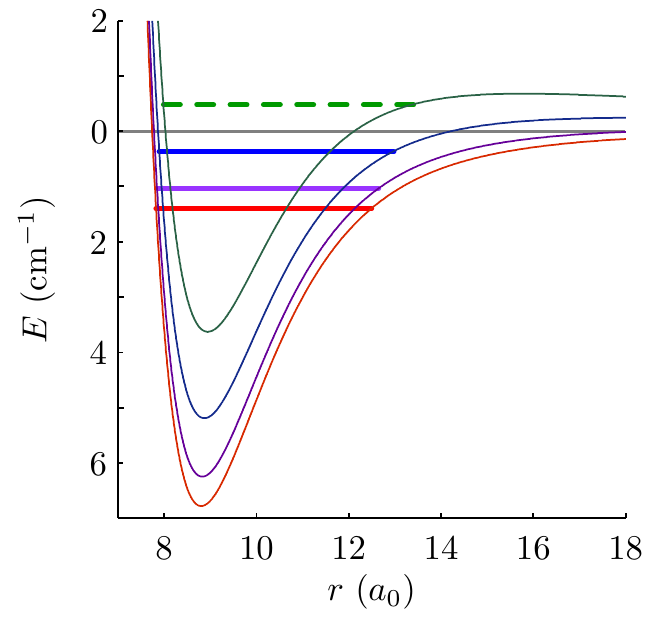}}\\
       (a)
  \end{minipage}
  \begin{minipage}[t]{2.28in}
       {\includegraphics[width=2.3in]{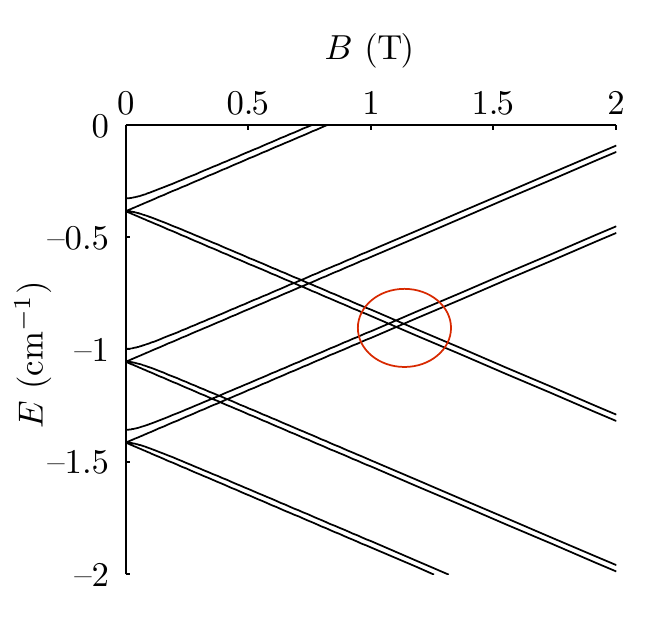}}\\
       (b)
  \end{minipage}
  \begin{minipage}[t]{2.40in}
       {\includegraphics[width=2.4in]{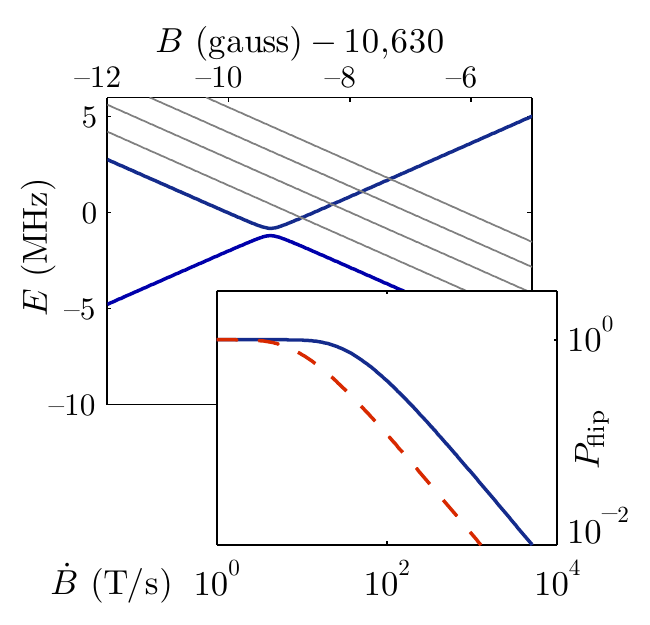}}
       (c)
  \end{minipage}
\caption{
(a) \textit{Ab initio} Potential energy for \aghethree\ $(^2\Sigma_+)$, showing the three bound $N=0,1,2$ molecular states (solid lines) and the quasibound $N=3$ state (dashed line).
(b) Zeeman level spectrum of \aghethree\ (circle indicates region of avoided crossings).
(c) Avoided level crossing between the $|N,m_N,m_S,m_{I\mathrm{Ag}},m_{I\mathrm{He}}\rangle = |0,0,1/2,-1/2,1/2\rangle$ and $|2,2,-1/2,-1/2,-1/2\rangle$ states.  The inset shows the spin flip probability as a function of $\dot{B}$ for $V=380$~kHz crossings (solid line) and 190~kHz crossings (dashed line).
}
\label{fig:agHeTotal}
\end{figure*}

To analyze the trap loss of \aghethree, we first calculated the potential energy curve to an expected accuracy of $10\%$, shown in \fig{fig:agHeTotal}~(a).  We employed the partially-spin restricted coupled cluster method with single, double and perturbative triple excitations (RCCSD(T))~\cite{knowles:combo} as implemented in the \textsc{molpro} suite of programs~\cite{MOLPRO}.
The potential supports one vibrational bound state, with $N = 0, 1, 2$ at energies $\epsilon_N = -1.40, -1.04, -0.37~\mathrm{cm}^{-1}$, and a quasibound $N=3$ state at $\epsilon_{3} = 0.48~\mathrm{cm}^{-1}$, with a calculated lifetime of 1 ns.
The molecular hyperfine constants were evaluated as a function of internuclear distance $r$ using a highly correlated density functional theory approach \cite{tscherbul:TBP}.  The values, averaged over the ground state molecular wavefunction, are $A_\mathrm{He} = -0.9~\mathrm{MHz}$ and $c = 1.04~\mathrm{MHz}$. The Zeeman spectrum of AgHe, consisting of identical hyperfine manifolds for each rotational level, is shown in \fig{fig:agHeTotal}~(b).

We estimate the \aghethree\ formation rate constant $K$ by assuming that pairs first form in the $N=3$ scattering resonance, and are subsequently quenched by rotational relaxation in additional collisions with He.
Using the resonant recombination model described in \cite{rbc:cpl:recombination:resonance}, we calculate values for $K$ between $0.8$ and $1.0\times 10^{-31}~\mathrm{cm}^6/\mathrm{s}$, for $T$ between $0.3$ and $0.7$~K, giving an equilibration time of $\leq 10~\mathrm{ms}$ for all $T$ and \nhe\ in the experiment.

To calculate the spin relaxation rate due to \aghecol\ collisions, we performed scattering calculations over the experimental temperatures and magnetic fields using the rigorous quantum formalism developed in \cite{tscherbul:alkali-He-collisions:pra2008,tscherbul:TBP}.  We set an upper bound on the spin-rotation constant $\gamma\leq 1.6~\mathrm{MHz}$ by scaling up the general formula from \cite{walker:pra1997:spinRotation} to match the observed relaxation at 0.7~K. We then averaged the relaxation rate coefficients over the trap field and thermal collision energy distribution \cite{brahms:thesis,hasted:collisions} to produce the dashed curve shown in \fig{fig:agHeRate}.

The calculated rate constant shows only a weak temperature dependence and exhibits marked disagreement with the experimental data below 0.6~K, suggesting that the observed loss rates \textit{cannot} be explained by atom-atom collisions.   We have also considered the \aghehecol\ collisional relaxation mechanism and, as expected from the estimates above, find it far too small to explain the experimental results (see \cite{tscherbul:TBP}).
\\
\\
\noindent\textit{Adiabatic transitions}
We finally consider the possibility of adiabatic transitions resulting from avoided crossings between the low-field-seeking $N=0$ and high-field-seeking $N=2$ Zeeman levels, shown in \fig{fig:agHeTotal}~(c). At a magnetic field of 1.06~T, the anisotropic hyperfine interaction couples these levels, causing avoided crossings \cite{krems:ybf:pra2007} with splittings ranging from 155 to 380~kHz. Trapped AgHe molecules crossing this magnetic field can experience a spin flip, causing the molecule to be expelled from the magnetic trap. We model this process as a Landau-Zener type adiabatic transition with probability
\begin{equation}
p_\text{flip}=1-\exp \left ( -\frac{\pi V^2}{\hbar\,\mu_B\, \vec{v}\cdot\vec{\nabla} B} \right ),
\end{equation}
which depends on the coupling strength $V$, the molecular translational velocity $\vec{v}$, and the magnetic field gradient~$\vec{\nabla}B$.  $p_\text{LZ}$ versus $\dot{B} = |\vec{v}\cdot\vec{\nabla} B|$ is shown in \fig{fig:agHeTotal}~(c).  

To understand how these transitions lead to an anomalous loss of Ag atoms, we use a direct simulation Monte Carlo approach \cite{oran:arfm:dsmc,brahms:thesis}.  For each temperature, 40,000 trapped Ag and \aghethree are simulated under the effects of the magnetic trapping field and hard-sphere collisions with the \hethree\ gas, with elastic collision probabilities calculated using the \textit{ab initio} \aghe\ potential energy in Fig.~\ref{fig:agHeTotal}~(a).  Each collision has a probability to cause molecular formation or dissociation according to the rate coefficients $K$ and $D=K/\kappa$, respectively.  We fit the results of the simulation to the measured spin change rate at 380~mK, extracting a value for $\epsilon_0$ of $-1.53\pm 0.08~\mathrm{cm}^{-1}$, indicating a theoretical underestimation of the potential depth on the order of $10\%$.  Using this fit potential, we then simulated the spin change rate vs.~temperature, finding agreement to the data over two orders of magnitude without further fitting.

We note that under typical experimental conditions, $\nx,\:\nxhe\ll \nhe$, and the formation rate of $X_2$ dimers is negligible.  Moreover, the experimental temperature far exceeds the 1~mK binding energy of $^4$He$_2$, ruling out the possibility for the formation of He dimers, and $\nhe$ is too low to produce the large He clusters that are found in supersonic expansions \cite{scoles:jcp:hebeams}.  For many species $\xhe_{p>1}$ clusters may also form.  The equilibrium density of these clusters is set by thermodynamics and can be made negligible by raising the system temperature.

In summary,\added[TW]{ our kinetic model for vdW pair formation and relaxation shows} that weakly bound atom-He heteronuclear molecules can form in copious quantities in buffer gas cooling experiments at sub-Kelvin temperatures. By performing quantum calculations of \aghecol\ interactions and dynamics, and identifying the trap loss mechanism for vdW molecules, we have explained the marked, heretofore unexplained, thermal dependence of buffer gas trapped Ag spin relaxation.

The molecules we predict will form in buffer gas experiments are trappable, and are slightly polar, with dipole moments $\sim 0.02~\mathrm{D}$.  It may be possible to prevent dissociation of the molecules by quickly reducing the He density---cryogenic valves can remove He in 40~ms \cite{harris:bgSmallMu}.  Were this performed with 300~mK Ag at a density of $10^{13}~\cmthree$, chemical equilibrium would be lifted at $\nhe < 10^{15}~\cmthree$, yielding a trapped sample of \aghethree\ at a density of $7\times 10^{10}~\cmthree$.

We wish to acknowledge stimulating discussions with A.~A.~Buchachenko and Y.~S.~Au.  We are grateful for grants from the DOE Office of Basic Energy Science and NSF to the Harvard-MIT CUA, ITAMP at the Harvard~/ Smithsonian Center for Astrophysics, the University of Wisconsin, and Millard Alexander at UMD (NSF CHE-0848110).

\bibliographystyle{apsrev}
\bibliography{AgHeDimer}

\end{document}